\documentclass[aps,prl,amsmath,amssymb,floatfix,twocolumn,amsmath,superscriptaddress,twocolumn,nofootinbib,tighten,letterpaper]{revtex4}
\usepackage{multirow}
\usepackage{bbold}
\usepackage{subfigure}
\usepackage{color}
\usepackage{mathrsfs}
\usepackage{hyperref}
\usepackage[normalem]{ulem}
\usepackage{bm}

\usepackage{amsfonts, relsize, color}
\usepackage{graphics}
\usepackage{graphicx}
\usepackage{subfigure}
\usepackage{hyperref}
\usepackage{color}
\usepackage{comment}
\usepackage{amsmath}
\usepackage{comment}

\usepackage{textgreek}

\newcommand{\ostar}{\mathbin{\mathpalette\make@circled\star}}

\renewcommand\vec[1]{\ensuremath\mathrm{\mathbf{#1}}}

\begin{document}
\title{\bf Anomalous and normal dislocation modes in Floquet topological insulators}

\author{Tanay Nag}
\affiliation{SISSA, via Bonomea 265, 34136 Trieste, Italy}
\affiliation{Institute f\" ur Theorie der Statistischen Physik, RWTH Aachen University, 52056 Aachen, Germany}

\author{Bitan Roy}~\thanks{Corresponding author: bitan.roy@lehigh.edu}
\affiliation{Department of Physics, Lehigh University, Bethlehem, Pennsylvania, 18015, USA}

\date{\today}
\begin{abstract}
\end{abstract}

\maketitle

\noindent 
{\bf Abstract}\\
{\bf Electronic bands featuring nontrivial bulk topological invariant manifest through robust gapless modes at the boundaries, e.g., edges and surfaces. As such this bulk-boundary correspondence is also operative in driven quantum materials. For example, a suitable periodic drive can convert a trivial insulator into a Floquet topological insulator (FTI) that accommodates nondissipative dynamic gapless modes at the interfaces with vacuum. Here we theoretically demonstrate that dislocations, ubiquitous lattice defects in crystals, can probe FTIs as well as unconventional $\pi$-trivial insulator in the bulk of driven quantum systems by supporting normal and anomalous modes, localized near the defect core. Respectively, normal and anomalous dislocation modes reside at the Floquet zone center and boundaries. We exemplify these outcomes specifically for two-dimensional (2D) Floquet Chern insulator and $p_x+ip_y$ superconductor, where the dislocation modes are respectively constituted by charged and neutral Majorana fermions. Our findings should be therefore instrumental in probing Floquet topological phases in the state-of-the-art experiments in driven quantum crystals, cold atomic setups, and photonic and phononic metamaterials through bulk topological lattice defects.}       

\begin{figure}[t!]
\includegraphics[width=0.75\linewidth]{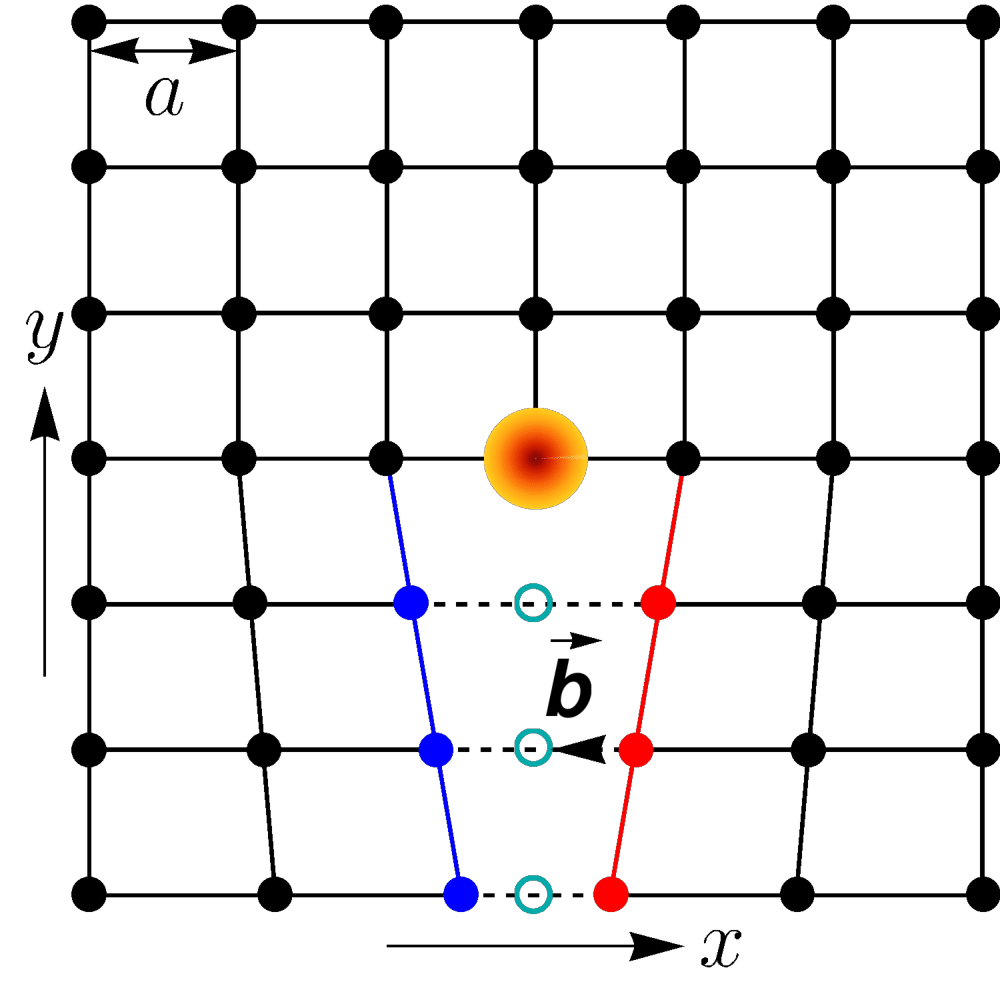}
\caption{{\bf Construction of a single edge dislocation}. Volterra cut-and-glue procedure on a square lattice by removing a line of atoms (open cyan circles), ending at the center or core of the dislocation (orange circle) and subsequently reconnecting (represented by the dashed lines) the sites (red and blue dots) living on the edges (red and blue lines) across it. The corresponding Burgers vector is ${\bf b}= a {\bf e}_x$.   
}~\label{fig:dislocationsetup}
\end{figure}

\begin{figure}[t!]
\includegraphics[width=0.95\linewidth]{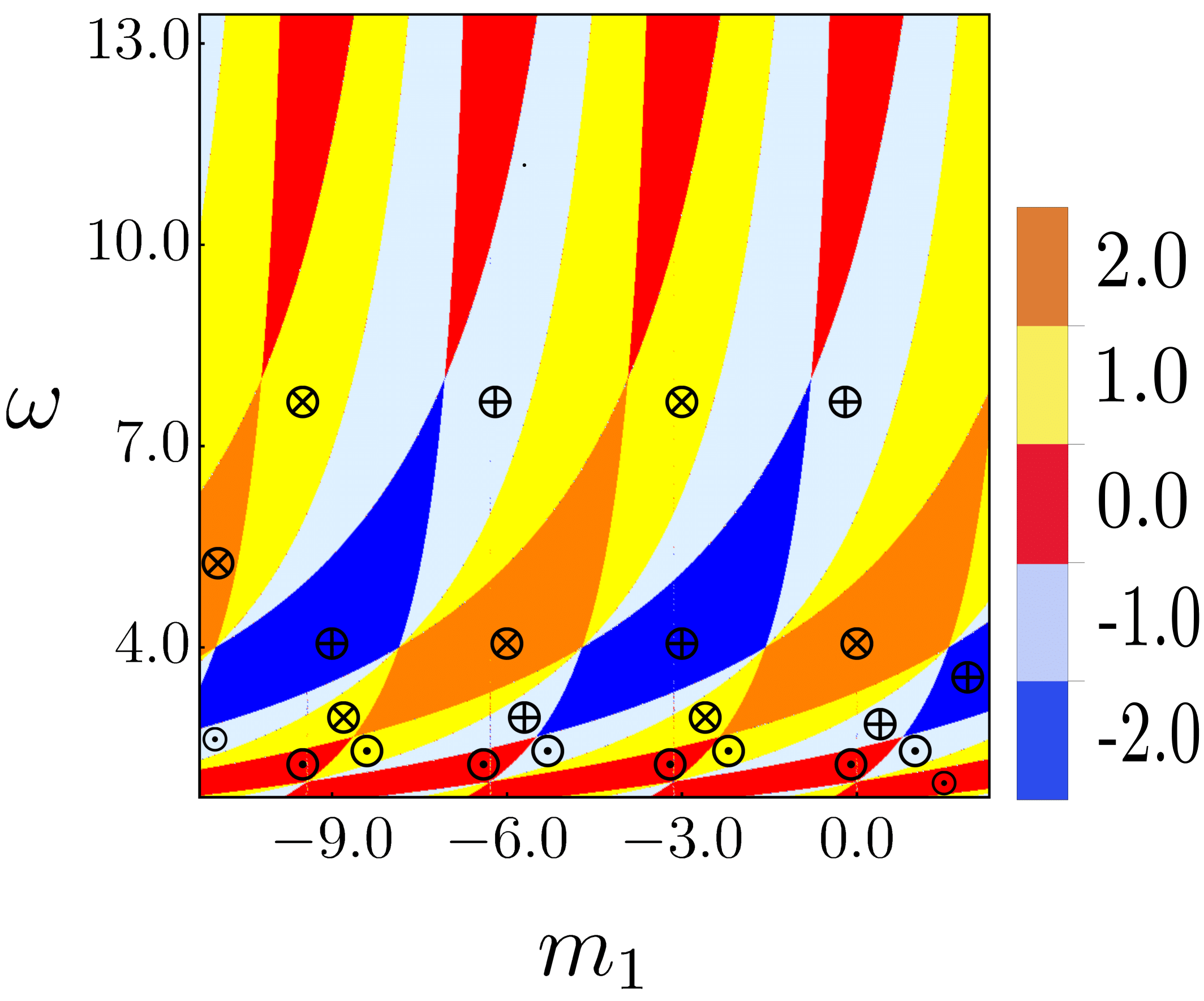}
\caption{{\bf Global phase diagram of a periodically driven time-reversal symmetry breaking insulator.} Here we set $t_1=t_0=1$ and $m_0=3$ [Eq.~(\ref{eq:statHamilt})], such that the static system is a trivial insulator, and $m_1$ and $\omega$ are respectively the amplitude and frequency of the external periodic drive [Eq.~(\ref{eq:perturbation_k})]. In the high frequency regime ($\omega \gtrsim 8$), the system only supports Floquet topological insulators (FTIs) with the Chern number $C=\pm 1$ and trivial ones with $C=0$ [Eq.~(\ref{eq:chernnumber})]. By contrast, in the intermediate and low frequency regimes ($\omega \lesssim 8$), the system in addition accommodates FTIs with $C=\pm 2$ and the $\pi$-trivial insulator with $C=0$. FTIs supporting normal (anomalous) dislocation modes at quasienergy $\mu = 0$ ($\pm \omega/2$) at the Flqouet zone center (zone boundary) are marked by $\otimes$ ($\oplus$). Whereas, Floquet insulators simultaneously supporting both normal and anomalous dislocation modes, such as the $\pi$-trivial insulator, are identified by $\odot$. Here we quote the Chern number of the conduction band, while that for the valence band is exactly the opposite.    
}~\label{fig:phasediagram}
\end{figure}

\begin{figure}
\includegraphics[width=1.0\linewidth]{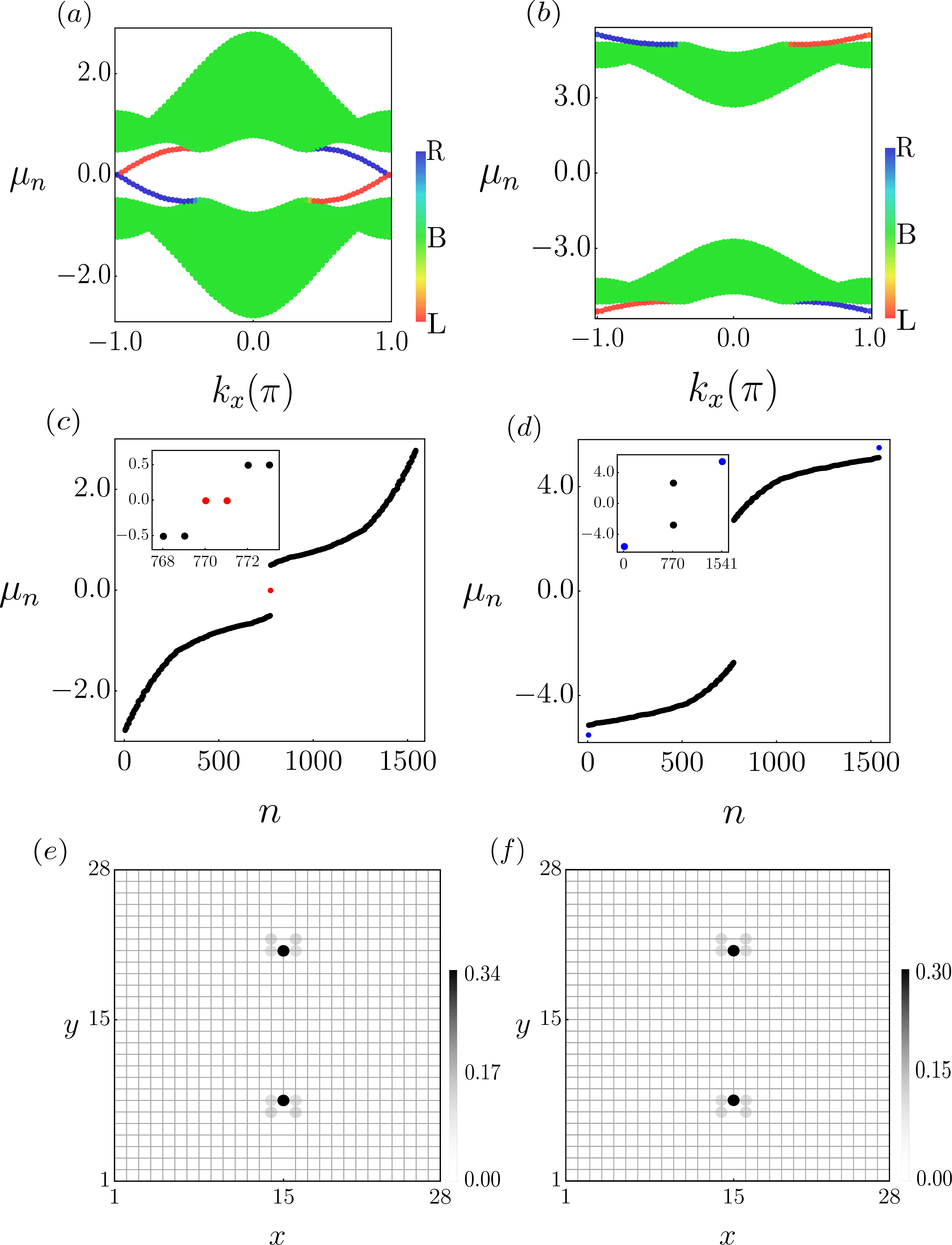}
\caption{{\bf Translationally active Floquet insulators in the high frequency regime.} Quasienergy ($\mu_n$) spectra in a semi-infinite system of linear dimension $L=100$ in the $y$ direction for (a) drive frequency $\omega=12.9$ and amplitude $m_1=-1.84$, yielding Chern number $C=+1$, and (b) $\omega=11.0$, $m_1=-5.3$, yielding $C=-1$. Blue and red [green] states are respectively localized on the right (R) and left (L) edges [in the bulk (B)], confirming that Floquet-Bloch band inversion occurs at $k_x=\pi$ at the Floquet zone center in (a) and zone boundary in (b). (c) Normal (red) and (d) anomalous (blue) dislocation modes at quasienergies $\mu=0$ and $\pm \omega/2$, respectively, well separated from bulk states (black), in a periodic system with a dislocation-antidislocation pair with Burgers vectors ${\bf b}=\pm a {\bf e}_x$. Local density of states for (e) normal and (f) anomalous dislocation modes.
}~\label{fig:highfreqC1}
\end{figure}

\begin{figure}[t!]
\includegraphics[width=1.0\linewidth]{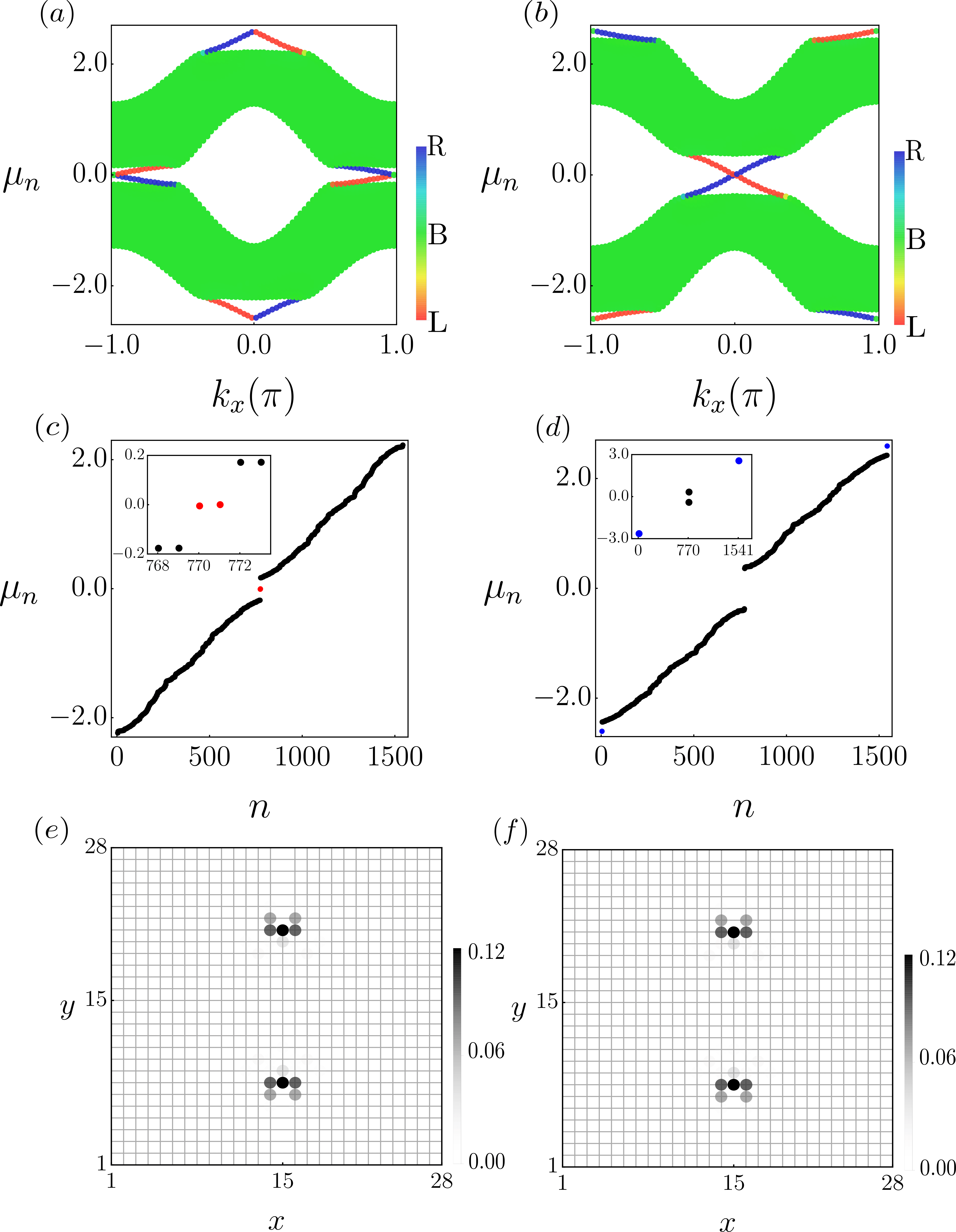}
\caption{{\bf Translationally active Floquet topological insulators (FTIs) in the intermediate frequency regime.}
Edge modes of FTIs with (a) Chern number $C=2$ for drive amplitude $m_1=-5.17$ and drive frequency $\omega=5.2$, and (b) $C=-2$ for $m_1=-8.32$ and $\omega=5.2$, in a semi-infinite system. The Floquet-Bloch band inversion occurs at $k_x=\pi$ ($0$) at (a) Floquet zone center (boundary) and (b) Floquet zone boundary (center). Respectively, these two phases support (c) normal (red) and (d) anomalous (blue) modes in a periodic system with a dislocation-antidislocation pair with Burgers vectors ${\bf b}=\pm a {\bf e}_x$. The local density of states for (e) normal and (f) anomalous dislocations modes. 
}~\label{fig:mediumfreqC2m2}
\end{figure}

\begin{figure}[t!]
\includegraphics[width=1.0\linewidth]{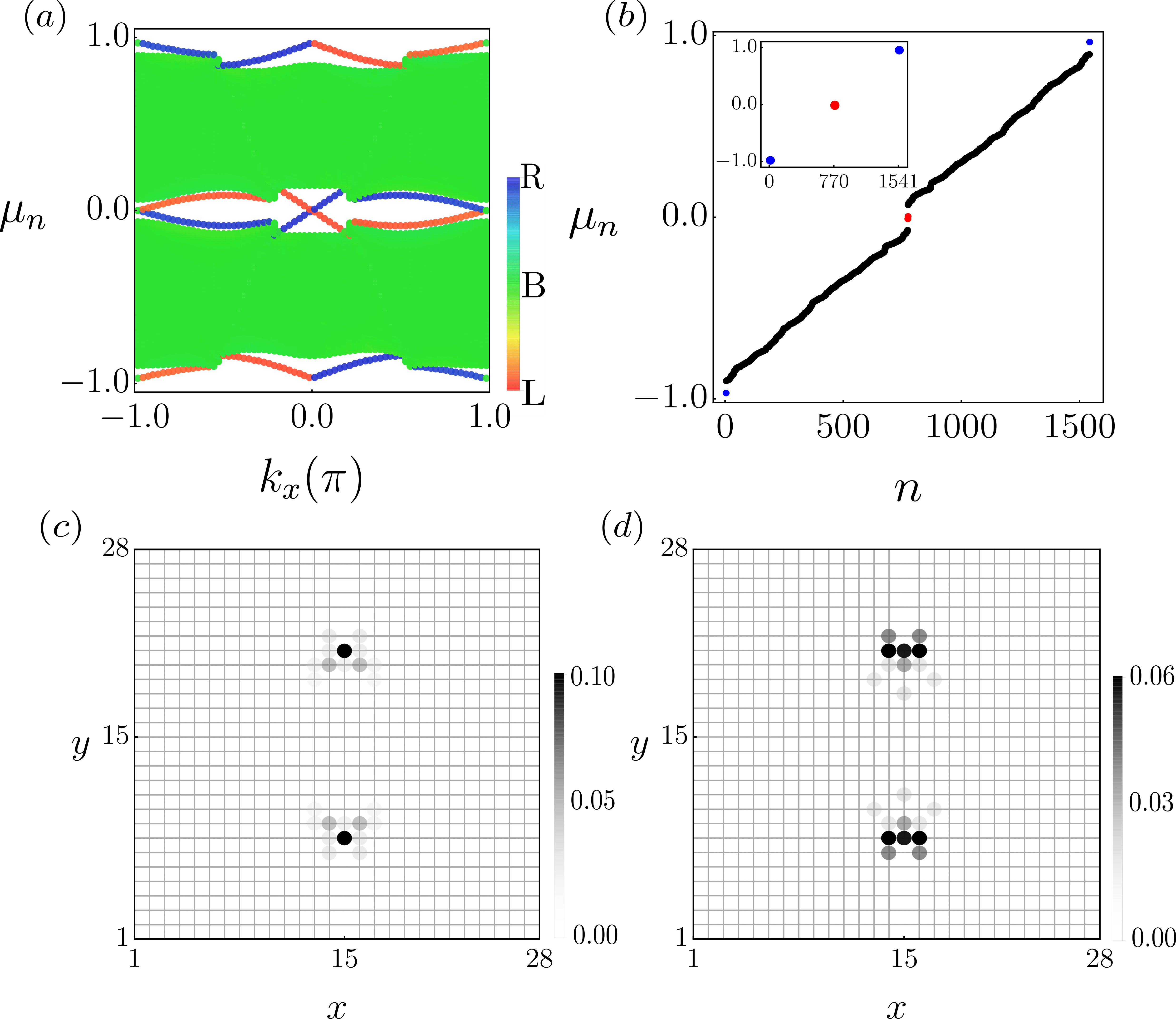}
\caption{{\bf Translationally active $\pi$ trivial insulator.} (a) Edge modes of a $\pi$-trivial insulator in a semi-infinite system at a low drive frequency $\omega=1.94$ and for drive amplitude $m_1=-4.5$, supporting band inversions at $k_x=\pi$ and $0$ at both Floquet zone center and boundary. (b) Normal (red) and anomalous (blue) modes bound to an edge dislocation-antidislocation pair. The corresponding local density of states are shown in (c) and (d), respectively. 
}~\label{fig:mediumfreqpitrivial}
\end{figure}

\noindent 
{\bf Introduction}\\
Quantum electronic materials can be classified into two broad categories: topological and trivial. In the former family, electronic wavefunctions in the bulk of the system feature nontrivial winding, resulting in gapless metallic states at an interface with vacuum~\cite{hasan-kane:RMP, qi-zhang:RMP}. This so-called the bulk-boundary correspondence beyond the territory of static systems also extends to the world of driven quantum materials~\cite{oka-aoki, galitski, berg-levin, zoller-budich, eckardt, takashi, berg-demler, moessner, dsenakdutta, JunHingAn, seradjeh, cooper, roy-harper, nag-juricic-roy}. But, due to the nontrivial role of the time dimension the bulk-boundary correspondence in driven systems is more subtle. For example, a static featureless electronic wavefunction subject to a suitable periodic drive can acquire nontrivial winding in the time direction, and due to the time translational symmetry the emergent dynamic boundary modes in the resulting Floquet topological phases are nondissipative in nature. Thus far however identification of dynamic topological matters somewhat exclusively relied on their boundary modes at the edges and surfaces~\cite{gedik, alu, azameit, Exp-2, Exp-4, Exp-5, Exp-MM-1, Exp-MM-3, Exp-MM-4}. Here we theoretically demonstrate that bulk topological lattice defects, namely dislocation [Fig.~\ref{fig:dislocationsetup}], can be instrumental in probing Floquet topological phases [Fig.~\ref{fig:phasediagram}], as they accommodate robust nondissipative modes, localized near the dislocation core [Figs.~\ref{fig:highfreqC1}-\ref{fig:mediumfreqpitrivial}], detectable in tunneling spectroscopy measurements, for example.

%%%%%%%%%%%%%%%%%%%%%%%%%%%%%%%%%%%%%%%%%%%%%%%%%%%
%%%%%%%%%%%%%%%%%%%%%%%%%%%%%%%%%%%%%%%%%%%%%%%%%%%
%%%%%%%%%%%%%%%%%%%%%%%%%%%%%%%%%%%%%%%%%%%%%%%%%%%
%%%%%%%%%%%%%%%%%%%%%%%%%%%%%%%%%%%%%%%%%%%%%%%%%%%
%%%%%%%%%%%%%%%%%%%%%%%%%%%%%%%%%%%%%%%%%%%%%%%%%%%
%%%%%%%%%%%%%%%%%%%%%%%%%%%%%%%%%%%%%%%%%%%%%%%%%%%
%%%%%%%%%%%%%%%%%%%%%%%%%%%%%%%%%%%%%%%%%%%%%%%%%%%
%%%%%%%%%%%%%%%%%%%%%%%%%%%%%%%%%%%%%%%%%%%%%%%%%%%
%%%%%%%%%%%%%%%%%%%%%%%%%%%%%%%%%%%%%%%%%%%%%%%%%%%
%%%%%%%%%%%%%%%%%%%%%%%%%%%%%%%%%%%%%%%%%%%%%%%%%%%

Dislocation in a 2D lattice is created by removing a line of atoms, ending at a site, known as its center or core, and subsequently joining the sites across the missing line of atoms, such that the translational symmetry is restored everywhere in the system, except near the core of the defect, see Fig.~\ref{fig:dislocationsetup}. As a result any closed loop around the dislocation center exhibits a missing translation by the Burgers vector ${\bf b}$, with ${\bf b}=a {\bf e}_x$ in Fig.~\ref{fig:dislocationsetup}, where $a$ is the lattice spacing. When an electron with momentum ${\bf K}$ encircles a dislocation it therefore picks up a hopping phase, given by $\exp[i \Phi_{\rm dis}]$, where $\Phi_{\rm dis}={\bf K} \cdot {\bf b}$ (modulo $2 \pi$). For topological phases of matter with the band inversion at momentum ${\bf K}_{\rm inv}$, this phase is given by $\Phi^{\rm top}_{\rm dis}={\bf K}_{\rm inv} \cdot {\bf b}$ (modulo $2 \pi$)~\cite{ran-zhang-vishwanath}. In a 2D topological insulator (electrical or thermal), introduction of a dislocation via reconnecting the sites across the line of removed atoms, therefore causes a level repulsion between the modes living on the edges across it (blue and red lines in Fig.~\ref{fig:dislocationsetup}). With the band inversion at a finite momentum of the Brillouin zone (BZ), as in the case of the $M$ phase with the band inversion at the $M=(1,1)\pi/a$ point, such a level repulsion between the edge modes is captured by a domain wall mass. It originates from the nontrivial $\pi$ hopping phase across the missing line of atoms (since $\Phi^{\rm top}_{\rm dis}=\pi$). The dislocation then binds topological zero-energy modes in the close vicinity of its core~\cite{ran-zhang-vishwanath, teo-kane, juricic-PRL, nagaosa, juricic-natphys, hughes-yao-qi, you-cho-hughes, roy-juricic-dislocation}, following the spirit of the Jackiw-Rebbi mechanism~\cite{jackiw-rebbi}. The $M$ phase thereby stands as an example of \emph{translationally active} topological insulator, as it features finite momentum band inversion at a non-$\Gamma$ point in the BZ, with ${\bf K}_{\rm inv}=(\pi,\pi)/a$ and supports robust zero-energy mode at the dislocation core, realized by breaking the translational symmetry in the bulk of the system as $\Phi^{\rm top}_{\rm dis}=\pi$ therein when ${\bf b}=a {\bf e}_x$, for example~\cite{juricic-PRL, juricic-natphys}. However, the role of dislocation lattice defects within the landscape of dynamic topological phases remained unexplored so far.

Here we demonstrate applicability of this general protocol in periodically driven FTIs. In particular, we show that dislocation defects can unveil 2D translationally active FTIs [Fig.~\ref{fig:phasediagram}], featuring finite momentum Floquet-Bloch band inversion, such as at the $M$ point, by localizing topologically robust nondissipative modes at its core. Otherwise, dislocation modes can be found either at the Floquet zone center (ZC) with vanishing quasienergy $\mu=0$ [Figs.~\ref{fig:highfreqC1}(c),~\ref{fig:highfreqC1}(e),~\ref{fig:mediumfreqC2m2}(c),~\ref{fig:mediumfreqC2m2}(e)], the normal dislocation modes or at the zone boundary (ZB) with $\mu=\pm \omega/2$ [Figs.~\ref{fig:highfreqC1}(d),~\ref{fig:highfreqC1}(f),~\ref{fig:mediumfreqC2m2}(d),~\ref{fig:mediumfreqC2m2}(f)], where $\omega$ is the frequency of the external periodic drive. The later ones are named anomalous dislocation modes as they lack any counterpart in static systems, where dislocation modes are found only at zero energy. Even in driven systems, the dislocation modes appear as topologically robust midgap states resulting from the hybridization between counter-propagating one-dimensional, otherwise normal or anomalous chiral edge modes that cross each other at quasienergy $\mu=0$ [Fig.~\ref{fig:highfreqC1}(a)] or $\pm \omega/2$ [Fig.~\ref{fig:highfreqC1}(b)] at a finite momentum, according to the ${\bf K} \cdot {\bf b}$ rule. The resulting normal (anomalous) dislocation modes therefore appear as the midgap states exactly at the quasienergy where the counter-propagating normal (anomalous) chiral edge modes cross each other at $k_x=\pi$, following the Jackiw-Rebbi mechanism. Although all the states supporting chiral edge mode at the Floquet ZB at $k_x=\pi$ or $0$ are anomalous~\cite{berg-levin, eckardt}, only the formers by virtue of finite momentum Floquet Bloch-band inversion accommodate anomalous dislocation modes.

%%%%%%%%%%%%%%%%%%%%%%%%%%%%%%%%%%%%%%%%%%%%%%%%%%%
%%%%%%%%%%%%%%%%%%%%%%%%%%%%%%%%%%%%%%%%%%%%%%%%%%%
%%%%%%%%%%%%%%%%%%%%%%%%%%%%%%%%%%%%%%%%%%%%%%%%%%%
%%%%%%%%%%%%%%%%%%%%%%%%%%%%%%%%%%%%%%%%%%%%%%%%%%%
%%%%%%%%%%%%%%%%%%%%%%%%%%%%%%%%%%%%%%%%%%%%%%%%%%%
%%%%%%%%%%%%%%%%%%%%%%%%%%%%%%%%%%%%%%%%%%%%%%%%%%%
%%%%%%%%%%%%%%%%%%%%%%%%%%%%%%%%%%%%%%%%%%%%%%%%%%%
%%%%%%%%%%%%%%%%%%%%%%%%%%%%%%%%%%%%%%%%%%%%%%%%%%%
%%%%%%%%%%%%%%%%%%%%%%%%%%%%%%%%%%%%%%%%%%%%%%%%%%%
%%%%%%%%%%%%%%%%%%%%%%%%%%%%%%%%%%%%%%%%%%%%%%%%%%%

More intriguingly, dislocations allow us to identify an unconventional phase, which we coin as the \emph{$\pi$-trivial insulator}, that despite possessing net zero topological invariant simultaneously supports normal and anomalous dislocation modes [Fig.~\ref{fig:mediumfreqpitrivial}]. Mixing between coexisting normal and anomalous dislocation modes are forbidden as they are separated by quasienergy $\omega/2$. We exemplify these outcomes for 2D Floquet Chern insulator and $p_x+ip_y$ superconductor for spinless or spin polarized fermions, respectively describing electrical and thermal insulators. While the former system accommodates charged dislocation modes, nondissipative localized Majorana fermions appear at the dislocation core of a translationally active Floquet $p_x+ip_y$ superconductor. Finally, we show that dislocations, besides being the unique probe of translationally active FTIs, accommodating Floquet-Bloch band inversion at finite momentum and localized normal and/or anomalous modes around its core, can play a pivotal role in their classification, when cannot be distinguished by bulk topological invariants.

\noindent 
{\bf Results}\\
{\bf Model, topology and ${\bf K} \cdot {\bf b}$ rule}. To illustrate the general principle of probing FTIs through bulk dislocation defects, here we focus on the paradigmatic examples of 2D time-reversal symmetry breaking insulators, namely the Chern insulator and $p_x+ip_y$ superconductor, described by the static Hamiltonian 
\begin{equation}~\label{eq:statHamilt}
H= \sum_{\vec{k}} \Psi^\dagger_{\vec{k}} \; \hat{h} (\vec{k}) \; \Psi_{\vec{k}}, \:\:\: 
{\rm with} \:\:\:
\hat{h}(\vec{k})=\sum^3_{\alpha=0} \sigma_\alpha \; d_\alpha (\vec{k}).
\end{equation}  
The above model supports topological insulators with band inversion at $\Gamma$ and $M$ points, characterized by distinct topological invariants, while only the later one being translationally active. For a Chern insulator, the two-component spinor is defined as $\Psi^\top_{\vec{k}}=\left( c_{A,\vec{k}}, c_{B, \vec{k}} \right)$. Here $c_{X,\vec{k}}$ is the fermion annihilation operator on orbital $X=A,B$ with momentum $\vec{k}$. Hereafter we set the lattice spacing $a=1$. The Pauli matrices $\{ \sigma_\mu \}$ then operate on the orbital indices. By contrast, for a $p_x+ip_y$ superconductor of spinless or spin polarized fermions the spinor reads $\Psi^\top_{\vec{k}}=\left( c_{\vec{k}}, c^\star_{-\vec{k}} \right)$. Then $\{ \sigma_\mu \}$ operate on the Nambu or particle-hole index. As $d_0(\vec{k})$ does not play any role in the topology or out of equilibrium dynamics, throughout we set $d_0(\vec{k})=0$ from the outset. The remaining components of the $\vec{d}$-vector are chosen to be
\begin{equation}
\vec{d}(\vec{k})= \bigg( t_1 \sin (k_x), t_1 \sin (k_y), m_0 - t_0 \sum_{j=x,y} \cos (k_j) \bigg). \nonumber 
\end{equation}

This model then features both topological and trivial insulating phases, respectively for $|m_0/t_0|<2$ and $|m_0/t_0|>2$. Furthermore, within the topological regime there exist two distinct phases for (a) $0<m_0/t_0<2$ with the band inversion at the $\Gamma$ point of the BZ, also known as the $\Gamma$ phase, and (b) $-2<m_0/t_0<0$ with the band inversion at the $M$ point of the BZ, also known as the $M$ phase. In the $p_x+ip_y$ paired state, the pairing in these two phases takes place in the close vicinity of a Fermi surface, realized near the $\Gamma$ and $M$ points, respectively. The system is then in the weak coupling regime. By contrast, in the trivial paired state, pairing occurs in the absence of a Fermi surface, and the system is then in the strong coupling regime. Irrespective of these details all the phases are characterized by the integer topological invariant, the first Chern number 
\begin{equation}~\label{eq:chernnumber}
C=\frac{1}{4\pi} \int_{\rm BZ} \; d^2 \vec{k} \; \left( \partial_{k_x} \hat{\vec{d}}(\vec{k}) \times \partial_{k_y} \hat{\vec{d}}(\vec{k}) \right) \cdot \hat{\vec{d}}(\vec{k}),
\end{equation}  
where $\hat{\vec{d}}(\vec{k})=\vec{d}(\vec{k})/|\vec{d}(\vec{k})|$. The momentum integral is performed over the first BZ. While $C=0$ in a normal insulator, in the $\Gamma$ ($M$) phase $C=-1$ ($+1$). Both of them support topologically protected one-dimensional chiral edge modes, yielding quantized charge and thermal Hall conductivities, given by $\sigma_{xy}=C e^2/h$~\cite{TKNN, haldane} and $\kappa_{xy}=C \pi^2 k^2_B T /(3h)$ as $T \to 0$~\cite{read-green}, respectively, in a Chern insulator and $p_x+ip_y$ paired state.

%%%%%%%%%%%%%%%%%%%%%%%%%%%%%%%%%%%%%%%%%%%%%%%%%%%
%%%%%%%%%%%%%%%%%%%%%%%%%%%%%%%%%%%%%%%%%%%%%%%%%%%
%%%%%%%%%%%%%%%%%%%%%%%%%%%%%%%%%%%%%%%%%%%%%%%%%%%
%%%%%%%%%%%%%%%%%%%%%%%%%%%%%%%%%%%%%%%%%%%%%%%%%%%
%%%%%%%%%%%%%%%%%%%%%%%%%%%%%%%%%%%%%%%%%%%%%%%%%%%
%%%%%%%%%%%%%%%%%%%%%%%%%%%%%%%%%%%%%%%%%%%%%%%%%%%
%%%%%%%%%%%%%%%%%%%%%%%%%%%%%%%%%%%%%%%%%%%%%%%%%%%
%%%%%%%%%%%%%%%%%%%%%%%%%%%%%%%%%%%%%%%%%%%%%%%%%%%
%%%%%%%%%%%%%%%%%%%%%%%%%%%%%%%%%%%%%%%%%%%%%%%%%%%
%%%%%%%%%%%%%%%%%%%%%%%%%%%%%%%%%%%%%%%%%%%%%%%%%%%

Even though the edge modes are protected by the nontrivial bulk topological invariant, the $\Gamma$ and $M$ phases respond distinctly when the translational symmetry, on the other hand, is broken in the bulk by introducing a dislocation. The real space Hamiltonian in the presence of an edge dislocation is shown in Supplementary Note 1 of the Supplemental Information (SI). Since $\Phi^{\rm top}_{\rm dis}=\pi$ ($0$) in the $M$ ($\Gamma$) phase, only the $M$ phase supports topological modes at the dislocation core, according to the ${\bf K} \cdot {\bf b}$ rule~\cite{juricic-PRL, juricic-natphys}. The dislocation mode gets pinned at zero energy due to an antiunitary particle-hole symmetry, generated by $\Theta=\sigma_1 {\mathcal K}$, as $\{ \hat{h}(\vec{k}), \Theta \}=0$, where ${\mathcal K}$ is the complex conjugation~\cite{broy-antiunitary}. The band inversion momentum (${\bf K}_{\rm inv}$) can be recognized from the zero-energy edge modes in a semi-infinite system with $k_x$ as a good quantum number and open boundaries in the $y$ direction, for example, which in the $M$ ($\Gamma$) phase appear near $k_x=\pi$ ($0$). See Fig.~S1 of the SI. The topological dislocation modes this way manifest the bulk-boundary correspondence in a static translationally active phase. We also show that dislocation modes are robust against modified hopping amplitudes along the ``distorted" bonds across the line of missing atoms ending at its core. See Fig.~S2 of the SI. In the rest of the paper, we show that the ${\bf K} \cdot {\bf b}$ rule is instrumental in identifying FTIs, which may or may not have any analogues in the static system.

%%%%%%%%%%%%%%%%%%%%%%%%%%%%%%%%%%%%%%%%%%%%%%%%%%%
%%%%%%%%%%%%%%%%%%%%%%%%%%%%%%%%%%%%%%%%%%%%%%%%%%%
%%%%%%%%%%%%%%%%%%%%%%%%%%%%%%%%%%%%%%%%%%%%%%%%%%%
%%%%%%%%%%%%%%%%%%%%%%%%%%%%%%%%%%%%%%%%%%%%%%%%%%%
%%%%%%%%%%%%%%%%%%%%%%%%%%%%%%%%%%%%%%%%%%%%%%%%%%%
%%%%%%%%%%%%%%%%%%%%%%%%%%%%%%%%%%%%%%%%%%%%%%%%%%%
%%%%%%%%%%%%%%%%%%%%%%%%%%%%%%%%%%%%%%%%%%%%%%%%%%%
%%%%%%%%%%%%%%%%%%%%%%%%%%%%%%%%%%%%%%%%%%%%%%%%%%%
%%%%%%%%%%%%%%%%%%%%%%%%%%%%%%%%%%%%%%%%%%%%%%%%%%%
%%%%%%%%%%%%%%%%%%%%%%%%%%%%%%%%%%%%%%%%%%%%%%%%%%%

{\bf Floquet insulators and dynamic dislocation modes}. To establish the applicability of the ${\bf K} \cdot {\bf b}$ rule within the Floquet framework, we focus on a static trivial insulator. For the rest of the discussion we, therefore, set $m_0=3$ and $t_1=t_0=1$, and periodically drive this system by the on site staggered potential 
\begin{equation}~\label{eq:perturbation_k}
\hat{V}(t)=m_1 \sigma_3 \sum_{r=-\infty}^{\infty} \delta(t-r T),
\end{equation}
where $r$ is an integer, $m_1$ is the drive amplitude, and $T$ denotes its period. The corresponding drive frequency is $\omega=2\pi/T$. For $p_x+ip_y$ pairing the above drive corresponds to a periodic modulation of the onsite chemical potential, and the resulting dislocation modes are constituted by localized Majorana fermions. The Floquet operator after a single kick in the momentum space is 
\begin{eqnarray}~\label{eq:floquetoperator}
U(\vec{k}, T) = {\rm TO} \left( \exp\left[ -i \int^T_0 \left[ \hat{h}(\vec{k}) + \hat{V}(t) \right] dt \right] \right), 
\end{eqnarray}
where `TO' stands for the time ordered product. The Floquet operator also satisfies the antiunitary particle-hole symmetry as $\Theta^{-1}U(\vec{k},T) \Theta=U(\vec{k},T)$. The associated effective Floquet Hamiltonian $\hat{h}_{\rm Flq}(\vec{k})=i \ln (U(\vec{k},T))/T$ takes the form $\hat{h}_{\rm Flq}(\vec{k})=\vec{d}_{\rm Flq}(\vec{k}) \cdot {\boldsymbol \sigma}$. The explicit form of the $\vec{d}_{\rm Flq}(\vec{k})$-vector in terms of the hopping and mass parameters is somewhat lengthy, which we show in Supplementary Note 2. In terms of the components of $\hat{h}_{\rm Flq}(\vec{k})$ we then compute the first Chern number ($C$) after taking $\vec{d}(\vec{k}) \to \vec{d}_{\rm Flq}(\vec{k})$ in Eq.~(\ref{eq:chernnumber}), and construct the global phase diagram of time-reversal symmetry breaking Floquet insulators, shown in Fig.~\ref{fig:phasediagram}.

To illustrate the response of FTIs to the bulk dislocation defects, first we diagonalize the Floquet operator from Eq.~(\ref{eq:floquetoperator}) in a semi-infinite system with $k_x$ as a conserved quantity and open boundaries in the $y$ direction [Figs.~\ref{fig:highfreqC1}(a),~\ref{fig:highfreqC1}(b),~\ref{fig:mediumfreqC2m2}(a),~\ref{fig:mediumfreqC2m2}(b),~\ref{fig:mediumfreqpitrivial}(a)]. It provides information regarding ${\bf K}_{\rm inv}$ of the Floquet-Bloch bands from the corresponding edge modes. Subsequently, we diagonalize the Floquet operator in a periodic system with a pair of dislocation-antidislocation. It reveals the existence of normal and/or anomalous dislocation modes [Figs.~\ref{fig:highfreqC1}(c),~\ref{fig:highfreqC1}(d),~\ref{fig:mediumfreqC2m2}(c),~\ref{fig:mediumfreqC2m2}(d),~\ref{fig:mediumfreqpitrivial}(b)], well separated from the bulk states, in translationally active FTIs, for which $\Phi^{\rm top}_{\rm dis}=\pm \pi$. We further anchor this outcome from the spatial profile of the local density states (LDoS) of these modes, displaying that they are strongly localized around the dislocation core [Figs.~\ref{fig:highfreqC1}(e),~\ref{fig:highfreqC1}(f),~\ref{fig:mediumfreqC2m2}(e),~\ref{fig:highfreqC1}(f),~\ref{fig:mediumfreqpitrivial}(c),~\ref{fig:mediumfreqpitrivial}(d)].

{\bf High frequency regime}.~In the high frequency regime ($\omega \gtrsim 8$) the system supports FTIs with $C=\pm 1$ and trivial one with $C=0$ [Fig.~\ref{fig:phasediagram}]. Only the Floquet insulators with $C=+1$ and $0$ can be qualitatively similar to the ones previously discussed for static system. The FTI with $C=+1$ featuring band inversion at the $M$ point at the Floquet ZC supports normal dislocation modes with $\mu \approx 0$ (due to the finite size effects). See Fig.~\ref{fig:highfreqC1}(a),(c),(e) and compare with Fig.~S1 of the SI. The trivial ones with $C=0$ are devoid of any dislocation mode, as in the static system. On the other hand, $C=-1$ FTIs can support edge modes at $k_x=\pi$, but at the Floquet ZB. Therefore, in such phases $C_{\rm ZC}=0$, but $C_{\rm ZB}=1$, yielding $C=C_{\rm ZC}-C_{\rm ZB}=-1$. Here $C_{\rm ZC}$ ($C_{\rm ZB}$) corresponds to the Chern number arising from the Floquet ZC (ZB)~\cite{berg-levin}. Therefore, the $C=-1$ FTI supports anomalous dislocation modes with quasienergies $\mu=\pm \omega/2$, according to the ${\bf K} \cdot {\bf b}$ rule, see Fig.~\ref{fig:highfreqC1}(b),(d),(f). By contrast, a static $C=-1$ insulator, with the band inversion at the $\Gamma$ point, does not support any dislocation mode. We discuss these cases in Supplementary Note 3 and the results are shown in Fig.~S3 of the SI.

Therefore, despite possessing same bulk topological invariant and equal number of edge modes, the $C=-1$ static Chern insulator and high-frequency FTI respond distinctly to dislocations. We also show that the Chern number ($C$) or the dynamic winding number ($W$)~\cite{berg-levin}, only yielding the total number of edge modes of a Floquet insulator in the entire Floquet zone, not ${\bf K}_{\rm inv}$, cannot identify translationally active FTIs. They can only be probed by dislocation modes, and FTIs with identical bulk invariants ($C$ and $W$) can respond distinctly to dislocations. See Table~S1 of the SI. Even though the normal and anomalous modes are predominantly localized at the dislocation cores, as the periodic drive introduces longer range hopping, the corresponding LDoS weakly spreads over few lattice sites away from the defect core, in comparison to its counterpart in a static system [Fig.~S1 of SI]. Still the overlap between the dislocation modes is quite negligible as long as the defect cores are sufficiently far apart. These features also persist in the medium and low frequency regimes, which we discuss next.

%%%%%%%%%%%%%%%%%%%%%%%%%%%%%%%%%%%%%%%%%%%%%%%%%%%
%%%%%%%%%%%%%%%%%%%%%%%%%%%%%%%%%%%%%%%%%%%%%%%%%%%
%%%%%%%%%%%%%%%%%%%%%%%%%%%%%%%%%%%%%%%%%%%%%%%%%%%
%%%%%%%%%%%%%%%%%%%%%%%%%%%%%%%%%%%%%%%%%%%%%%%%%%%
%%%%%%%%%%%%%%%%%%%%%%%%%%%%%%%%%%%%%%%%%%%%%%%%%%%
%%%%%%%%%%%%%%%%%%%%%%%%%%%%%%%%%%%%%%%%%%%%%%%%%%%
%%%%%%%%%%%%%%%%%%%%%%%%%%%%%%%%%%%%%%%%%%%%%%%%%%%
%%%%%%%%%%%%%%%%%%%%%%%%%%%%%%%%%%%%%%%%%%%%%%%%%%%
%%%%%%%%%%%%%%%%%%%%%%%%%%%%%%%%%%%%%%%%%%%%%%%%%%%
%%%%%%%%%%%%%%%%%%%%%%%%%%%%%%%%%%%%%%%%%%%%%%%%%%%

{\bf Medium and low frequency regimes}.~In the medium and low frequency regime, when the drive frequency is comparable to the bandwidth of the static system, the adjacent Floquet zones get even strongly coupled, which in turn gives rise to FTIs with no analogues in the static system. For example, in this regime we find FTIs with $C=\pm 2$ [Fig.~\ref{fig:phasediagram}]. The appearances of these two phases and the corresponding Chern number can be appreciated from their edge modes. The FTI with $C=2$ features edge modes at $k_x=\pi$ and $0$ at the Floquet ZC and ZB [Fig.~\ref{fig:mediumfreqC2m2}(a)], respectively, yielding $C_{\rm ZC}=+1$ and $C_{\rm ZB}=-1$, and therefore $C=C_{\rm ZC}-C_{\rm ZB}=2$. Concomitantly, this phase only supports normal dislocation modes with $\mu \approx 0$ [Figs.~\ref{fig:mediumfreqC2m2}(c),~\ref{fig:mediumfreqC2m2}(e)], as the finite momentum Floquet-Bloch band inversion occurs only at the Floquet ZC. On the other hand, FTI with $C=-2$ features edge modes at $k_x=0$ and $\pi$ at the Floquet ZC and ZB, respectively, and thereby yielding $C_{\rm ZC}=-1$ and $C_{\rm ZB}=+1$ [see Fig.~\ref{fig:mediumfreqC2m2}(b)]. As a result of the finite momentum band inversion only at the Floquet ZB, this phase supports anomalous dislocation modes at quasienergies $\mu=\pm \omega/2$ [Fig.~\ref{fig:mediumfreqC2m2}(d),~\ref{fig:mediumfreqC2m2}(f)], also in agreement with the ${\bf K} \cdot {\bf b}$ rule. Therefore, the FTIs with the Chern number $C=+2$ and $-2$ become translationally active at the Floquet ZC and ZB, respectively.

More intriguingly, at low frequency we also identify an insulating phase with $C=0$ [Fig.~\ref{fig:phasediagram}], which we name the $\pi$-trivial insulator, that features edge modes at $k_x=0$ and $\pi$ at both Floquet ZC and ZB [Fig.~\ref{fig:mediumfreqpitrivial}(a)]. Following the above rule, we find that in this phase $C=0$ as $C_{\rm ZC}=C_{\rm ZB}=0$. Due to the existence of the edge modes at $k_x=\pi$ at both Floquet ZC and ZB, this phase however simultaneously supports normal and anomalous dislocation modes [Fig.~\ref{fig:mediumfreqpitrivial}(b)], according to the ${\bf K} \cdot {\bf b}$ rule. These modes are also strongly localized at the dislocation cores [Figs.~\ref{fig:mediumfreqpitrivial}(c),(d)]. Therefore, the $\pi$-trivial insulator is translationally active at both Floquet ZC and ZB.

In addition, $C=\pm 1$ FTIs in the low frequency regime can support both normal and anomalous dislocation modes, in contrast to the situation in the high frequency regime [Fig.~\ref{fig:highfreqC1}]. We discuss these scenarios in Supplementary Note 4 and the results are displayed in Fig.~S4 of the SI. Furthermore, in the low frequency regime $C=+1$ or $-1$ FTIs despite possessing equal number of edge modes or the dynamic winding number, can be completely different depending on the Floquet-Bloch band inversion momentum in the Floquet zone. Consequently, they respond distinctly to dislocations. See Table~S1 of the SI.

The realizations of dislocation modes are, however, qualitatively insensitive to the particular choice of the drive protocol. We discuss two distinct drive protocols in Supplementary Note 5, and the associated dislocation modes are shown in Fig.~S5 and Fig.~S6 of the SI.

\noindent 
{\bf Discussions}\\
Considering the simplest realization of 2D Floquet insulators, here we show that dislocation lattice defects can be instrumental in unveiling a rich landscape of translationally active dynamic topological phases, supporting normal and anomalous dislocation modes around the defect core. This mechanism is also applicable for light-induced insulators~\cite{oka-aoki, moessner, gedik}, dynamic metamaterials~\cite{azameit, Exp-MM-1, Exp-MM-3, Exp-MM-4, alu, Exp-2}, as well as in cold atomic setups~\cite{Exp-4, Exp-5}. Normal and anomalous dislocation modes are equally germane to a driven $p_x+i p_y$ superconductor, where they are constituted by localized Majorana fermions.

Furthermore, the ${\bf K} \cdot {\bf b}$ rule should be applicable to three-dimensional Floquet phases, encoding their responses to edge dislocation, constructed by stacking its 2D counterpart~\cite{juricic3Ddislocation} and screw dislocations, where electron picks up a nontrivial phase as it hops between the neighboring layers through the ``slipping half-plane"~\cite{ran-zhang-vishwanath}. As we show that the requisite finite momentum Floquet-Bloch band inversion can in principle be engineered by tuning the amplitude and/or frequency of the external periodic drive. This general protocol therefore opens an experimentally viable route to probe translationally active Floquet phases through bulk topological lattice defects in topological crystals, featuring band inversion at the $\Gamma$ point or even in trivial systems with no band inversion at all in the static limit. Therefore, dislocations, standing as the unique probe for translationally active FTIs, hosting localized normal and anomalous modes around their core, along with the bulk invariants (such as the Chern and dynamic winding numbers~\cite{berg-levin}), provide a more complete classification of FTIs, especially when their bulk invariant are the same. See Table~S1 of the SI.

With a suitable drive protocol, it is also conceivable to realize more than one topological edge modes at $k_x=\pi$ at the Floquet ZC and/or ZB~\cite{gong2020}. Then according to the proposed ${\bf K} \cdot {\bf b}$ rule each copy of finite momentum chiral edge mode yields one dislocation mode at the corresponding quasienergy. See Supplementary Note 6 and Fig.~S7 of the SI. If, on the other hand, chiral edge modes appear at momentum (${\bf K}_{\rm inv}$) away from the BZ center ($\Gamma$ point) or boundary ($M$ point)~\cite{gong2018}, then robust dislocation modes can be found when the corresponding Burgers vector satisfies ${\bf K}_{\rm inv} \cdot {\bf b}=\Phi^{\rm top}_{\rm dis}=\pi$ (modulo $2 \pi$).

With the recent progress in probing static topological phases via lattice defects~\cite{edagawa, beidenkopf} and the existing experimental advancements to Floquet engineer topological phases~\cite{gedik, alu, azameit, Exp-2, Exp-4, Exp-5, Exp-MM-1, Exp-MM-3, Exp-MM-4}, we expect that the proposed dynamic dislocation modes can be observed experimentally in the near future. In addition, our proposal should be germane in driven metamaterials, e.g., in photonic~\cite{azameit, Exp-MM-1, Exp-MM-3, Exp-MM-4} and acoustic~\cite{alu, Exp-2} lattices, where FTIs has been observed, and lattice defects can be engineered and manipulated~\cite{photonic-dislocation-1, photonic-dislocation-2, photonic-disclination, mechanical-dislocation}. For example, in photonic~\cite{azameit, Exp-MM-1, Exp-MM-3, Exp-MM-4} and acoustic~\cite{alu, Exp-2} lattices dislocations can be created by removing a line of optical waveguides~\cite{photonic-dislocation-2} and magnetomechanical resonator~\cite{mechanical-dislocation} up to its core, respectively, and subsequently joining the ones across the missing line. The dislocation modes can be detected from LDoS, which in electronic systems can be measured by scanning tunneling microscope (STM)~\cite{beidenkopf}. While in photonic crystals LDoS can be obtained from two-point pump probe~\cite{photonic-dislocation-2} or reflection spectroscopy~\cite{photonic-disclination}, in acoustic lattices mechanical susceptibility at each magnetomechanical resonator yields LDoS~\cite{mechanical-dislocation}. Our proposed drive protocol for staggered on site potential can in principle be realized on optical lattices~\cite{goldmanPRX}, by changing the lattice depth, for example, which has also been used to tune the strength of on site Hubbard interaction across a quench~\cite{blochquench}.        

\noindent 
{\bf Methods}\\
To capture the response of Floquet insulators to dislocation lattice defects we pursue the following general approach. For a system, described by the Hamitlonian $\hat{h}(\vec{k})$ in the static limit and subject to a generic periodic drive $\hat{V}(t)$, such that $\hat{V}(t+T)=\hat{V}(t)$, where $T$ is the period of the drive, we construct the associated Floquet or the time evolution operator $U(\vec{k},T)$ at the stroboscopic time $t=T$, the explicit form of which is displayed in Eq.~(\ref{eq:floquetoperator}). First, we compute $U(k_x,T)$ in a semi-infinite system, with $k_x$ as good quantum number and open boundaries in the $y$ direction along which the linear dimension of the system is $L$, and typically $L=100$. This step is repeated for various drive amplitude and frequency. It allows us to identify Floquet states that possess counter propagating edge modes crossing each other at $k_x=\pi$ around the Floquet zone center and/or boundary. Such states feature Floquet Bloch band inversion at the $M$ point of the 2D Brillouin zone. Next, for the same set of parameter values, we compute the quasimodes of the Floquet operator in real space with an edge dislocation-antidislocation pair. By virtue of ${\bf K} \cdot {\bf b}=\pm \pi$, stemming from the interplay of the band inversion at the $M$ point and the Burgers vector of the dislocation ${\bf b}= \pm a{\bf e}_x$, these states support quasimodes at the Floquet zone center with quasienergy $\mu=0$ and/or zone boundary with quasienery $\mu=\pm \omega/2$, where $\omega=2 \pi/T$. To this end we numerically solve for the quasimodes $|\mu_n \rangle$ with quasienergies $\mu_n$ at the stroboscopic time $t=T$, satisfying $U(\vec{k},T)|\mu_n \rangle =\exp \left( i \mu_n T \right) |\mu_n \rangle$ in the aforementioned geometries. We follow this procedure for the drive protocol shown in Eq.~(\ref{eq:perturbation_k}) of the main manuscript, and the other drive protocols discussed in the Supplementary Note 5 and Supplementary Note 6.      

\noindent 
{\bf Data availability} \\
The data that support the plots within this paper and other findings of this study are available from the corresponding author and Tanay Nag (tanaynag23@gmail.com) upon reasonable request.

\noindent 
{\bf Acknowledgments} \\
Tanay Nag thanks MPIPKS, Dresden for the computation facilities. Bitan Roy was supported by the Startup grant from Lehigh University. We thank Vladimir Juri\v ci\' c and Andras L. Szab\' o for discussions. 

\noindent 
{\bf Author contributions} \\
Bitan Roy conceived and structured the project, and wrote the manuscript. Tanay Nag performed all the numerical calculations.  

\noindent 
{\bf Competing Interests} \\
The authors declare no competing interests.

%%%%%%%%%%%%%%%%%%%%%%%%%%%%%%%%%%%%%%%%%%%%%%%%%%%%%%%%%%%%
%%%%%%%%%%%%%%%%%%%%%%%%%%%%%%%%%%%%%%%%%%%%%%%%%%%%%%%%%%%%
%%%%%%%%%%%%%%%%%%%%%%%%%%%%%%%%%%%%%%%%%%%%%%%%%%%%%%%%%%%%
%%%%%%%%%%%%%%%%%%%%%%%%%%%%%%%%%%%%%%%%%%%%%%%%%%%%%%%%%%%%
%%%%%%%%%%%%%%%%%%%%%%%%%%%%%%%%%%%%%%%%%%%%%%%%%%%%%%%%%%%%

\end{document}